\documentclass[12pt,oneside,a4paper]{book} 

\usepackage{natbib}
\usepackage{times,amsmath,float}
\usepackage{amssymb}
\usepackage{graphicx}
\usepackage{subfigure}
\usepackage{amsfonts}
\usepackage{epsfig}
\usepackage{anysize}
\usepackage{xtab}
\usepackage{lscape}
\usepackage{geometry}
\usepackage{longtable}
\usepackage{appendix}
\usepackage{setspace}
\usepackage{epstopdf}
\usepackage{chngcntr}
\usepackage{booktabs}
\usepackage{multirow}
\usepackage{caption,booktabs,array}
\usepackage{float}
\usepackage{adjustbox}
\usepackage{url}
\usepackage{multicol}
\usepackage{placeins}
\usepackage{pdfpages}
\usepackage{multicol}
\usepackage{titling}

\usepackage{xcolor}
\usepackage{graphicx}

\usepackage{titlesec}
\titleformat{\chapter}[display]{\normalfont\normalsize\centering\upshape}{\MakeUppercase{\chaptertitlename}\ \thechapter}{8pt}{\normalsize}
 
  \titleformat{\section}
  {\normalfont\normalsize\centering\upshape}{}{1em}{}

  \titleformat{\subsection}
{\normalfont\normalsize\centering\upshape}{}{1em}{}

\titlespacing{\chapter}{0pt}{-50pt}{0.3cm}

\usepackage{fancyhdr}
\usepackage[justification=justified,singlelinecheck=false]{caption}
\usepackage[labelsep=period]{caption}

\marginsize{4cm}{2.5cm}{2.5cm}{2.5cm}

\counterwithout{figure}{chapter}
\counterwithout{table}{chapter}
\counterwithout{footnote}{chapter}
\counterwithout{equation}{chapter}


\begin{document}

 \setlength{\parindent}{2.1em}
\doublespacing 

\frontmatter

\begin{center}

\begin{center}

\vspace{7cm}
\LARGE Environmental Kuznets Curve \& Effectiveness of International Policies: Evidence from Cross Country Carbon Emission Analysis
\vspace{4cm}

\Large Elvan Ece Satıcı\\
\&\\
\Large Bayram Cakir
\vspace{4cm}

\Large May 2021
\end{center}
\pagebreak 

\chapter{ABSTRACT}


\end{center}

\indent In this article, we are presenting the relationship between the environmental pollution and the income level of the selected twenty-four countries.  We implemented a data based research analysis where, for each country, we analyzed the related data for fifty-six years, from 1960 to 2016, to assess the relationship between the carbon emission and the income level. \\
 \indent After performing the related data analysis for each country, we concluded whether the results for that country were in line with the Environmental Kuznets Curve (“EKC”) hypothesis. The EKC hypothesis suggests that the carbon emission per capita starts a declining trend when the country specific high-level of income is reached. The results of our data analyses show that the EKC hypothesis is valid for high-income countries and the declining trends of the carbon emission are clearly observed when the income level reaches a specific high enough level. On the other hand, for the non-high income countries our analysis results show that it is too early to make an assessment at this growth stage of their economies because they have not reached their related high-enough income per capita levels yet.  \\
 \indent Furthermore, we performed two more additional analysis on high-income countries.  First, we analyzed the related starting years of their carbon emission declining trends. The big variance in the starting years of the carbon emission declining trends show that the international policies are clearly ineffective in initiating the declining trend in carbon emission. In addition, for the high-income countries, we explained the differences in their carbon emission per capita levels in 2014 with their SGI indices and their dependence on high-carbon emission energy production.

\pagebreak
\singlespacing

\chapter{TABLE OF CONTENTS}

\vspace*{0.4cm}
\doublespacing
\noindent CHAPTER 1: INTRODUCTION \dotfill 1 \hspace{1cm}\\
\noindent CHAPTER 2: LITERATURE REVIEW \dotfill 5 \hspace{1cm}\\
\noindent CHAPTER 3: DATA \dotfill 8 \hspace{1cm}\\
\noindent CHAPTER 4: D.	ANALYSIS OF THE SELECTED COUNTRIES  \dotfill 10 \hspace{1cm}\\
\noindent CHAPTER 5: THE START OF THE CARBON EMISSION DECLINE \dotfill 10 \hspace{1cm}\\
\noindent CHAPTER 6: HIGH-INCOME COUNTRIES' CARBON EMISSION LEVELS \dotfill 10 \hspace{1cm}\\
\noindent CHAPTER 7: CONCLUSION \dotfill 25 \hspace{1cm}\\
\noindent REFERENCES \dotfill 27 \hspace{1cm}\\

\mainmatter
\chapter{INTRODUCTION}
\vspace*{0.4 cm}

\indent In this article, we are presenting the relationship between the environmental pollution and the income level of the selected twenty-four countries.  In assessing the relationship between the carbon emission and the income level, we implemented a data based research analysis where, for each country, we analyzed the related data for fifty-six years, from 1960 to 2016. \\
\indent In section B, we are explaining the Environmental Kuznets Curve hypothesis and we are providing a brief literature review on the topic. \\
\indent In section C, we are defining the data used in this research as the base of our analyses. \\
\indent In section D, we analyze each one of the selected twenty-four countries in order to understand the relationships between their income levels and their carbon emission levels. We classify these countries under two categories as “high-income countries” and “non-high income countries” based on the World Bank’s data on “GDP per capita”. The selected countries with GDP per capita levels of more than 20,000 USD are categorized as the “high-income” countries whereas the other countries with GDP per capita below 12,000 USD are referred to as the “non-high income” countries.  For each one of the selected twenty-four countries, income per capita and carbon emission per capita data for fifty-six years, ranging from 1960-2016, is used in the analysis to present the environmental pollution trend with respect to the income level.  At the end of the data analysis for each country, we conclude whether the carbon emission trend in the related country was in line with the Environmental Kuznets Curve (“EKC”) hypothesis. \\
\indent In section E, we concentrate on the high-income countries and analyze the starting year of the carbon emission per capita declining trend with respect to the related GDP per capita. If the international policies were effective in starting the carbon emission per capita declining trend in these high-income countries, we would expect to observe a convergence around the specific years when these policies were put in effect.  In addition to analyzing the effectiveness of the international policies, we also analyze the effect of the national efforts in the initiation of the carbon emission per capita decline trend by using the SGI scores of the related countries.  Briefly, SGI score states the country’s overall diligence on the environment with respect to the effectiveness of the policies and regulations enforced by their governments. \\
\indent In section F, we present further information on high-income countries and focus on their levels of carbon emission per capita in 2014. We explain the reasons behind the differences in carbon emission per capita levels in 2014 with the countries’ related SGI indices and their dependence on high-carbon emission energy production.

\clearpage

\chapter{LITERATURE REVIEW}
\vspace*{0.4 cm}
\doublespacing
\indent Due to the steady increase in the air pollution in the past decades, researches have started to focus on understanding the dynamics between the income levels of the countries and their environmental degradation levels.  In fact, the concept of “Environmental Kuznets curve” (“EKC”) was mentioned first in 1991 by Grossman and Krueger in their study of North American Free Trade Agreement (NAFTA) (Hervieux and Mahieu 2014). Grossman and Krueger stated in their paper titled “Environmental Impacts of a North American Free Trade Agreement” that “We find for two pollutants (sulfur dioxide and "smoke") that concentrations increase with per capita GDP at low levels of national income, but decrease with GDP growth at higher levels of income”  (1991). \\ 
\indent In 1992, the World Bank published its report titled “World Development Report 1992: Development and the Environment” which argued that economic growth would enable improved environmental conditions if policies and programs were put in place: \\
\indent The main message of this year's report is the need to integrate environmental \

considerations into development policymaking. The report argues that 

\ continued, and even accelerated, economic and human development is  

\ sustainable and can be consistent with improving environmental conditions, but 

\ that this will require major policy, program, and institutional shifts. A twofold 

\ strategy is required. First, the positive links between efficient income growth 

\ and the environment need to be aggressively exploited. Second, strong policies 

\ and institutions need to be put in place which cause decision makers to 

\ adopt less damaging forms of behavior. \

\ (World Bank 1992) \ \

\indent Grossman and Krueger named the Environmental Kuznets Curve (“EKC”) after its resemblance to Kuznet’s s (1955) inverted U-shaped relationship between income inequality and development (Dasgupta et al. 2002).  In their article titled “Confronting the Environmental Kuznets Curve”, Dasgupta, Laplante, Wang and Wheeler explained the EKC hypothesis as follows:\\
\indent  In the first stage of industrialization, pollution in the environmental \ 

\ Kuznets curve world grows rapidly because people are more interested in jobs 

\ and income than clean air and water, communities are too poor to pay for 

\ abatement, and environmental regulation is correspondingly weak. The balance 

\ shifts as income rises. Leading industrial sectors become cleaner, people 

\ value the environment more highly, and regulatory institutions become 

\ more effective. Along the curve, pollution levels off in the middle-income 

\ range and then falls toward pre-industrial levels in wealthy societies.

\ (Dasgupta et al. 2002)

\indent Following the publication of this World Bank report in 1992, many researchers have conducted numerous researches on this subject globally, using different techniques and various pollutants as variables.

\clearpage

\chapter{THE DATA}
\vspace*{0.4 cm}
\doublespacing
\indent In this article, Gross Domestic Product (“GDP”) per capita (constant 2010 USD) data is used as a reference to indicate the income level of each of the selected countries.  GDP per capita (constant 2010 USD) data for the selected countries is obtained from the World Bank database.  GDP per capita (constant 2010 USD) is defined as the gross domestic product of the related country divided by its midyear population. The GDP is calculated as the sum of gross value added by all resident producers in the economy plus any product taxes and minus any subsidies not included in the value of the products, without making any deductions for depreciation of fabricated assets or for depletion and degradation of natural resources (“GDP per capita (Constant 2010 USD)” 2021). GDP per capita data used in this article is “in constant 2010 U.S. dollars” to ensure comparability among the years in the related period, which is from 1960 to 2016.  In this article, the countries with GDP per capita above 20,000 USD in 2016 are considered as high-income countries whereas the countries with GDP per capita below 12,000 USD in 2016 are referred as “non-high income” countries. The threshold income levels used in this article for the categorization of the high-income and non-high income countries are aligned with the World Bank definitions, stating that high-income countries are the ones with Gross National Income per capita of 12,275 USD or more in 2010 (Beer and Prydz 2019).
Analogous to the data for GDP per capita, the carbon dioxide emissions data for the related period is also obtained from the World Bank database. The World Bank defines carbon emission per capita as the carbon dioxide emissions from the “burning of fossil fuels and the manufacture of cement,”  where the “carbon dioxide produced during consumption of solid, liquid and gas fuels and gas flaring” is also included (“CO2 Emissions (Metric Tons per Capita)”  2021). The data is available from 1960 to 2016 for most of the countries except for Germany where the data obtained starts from 1991, and France and Italy for which the data is only available until 2014. \ 

\indent The Environmental Report of the Sustainable Governance Indicators measures the effectiveness of the policies and regulations enforced by the governments. The qualitative measures addressed by this report are the ambition levels of the environmental targets the governments seek to fulfill, the impact of the implementation, and the integration of the policies in relevant sectors (Stiftung 2020). In this report, 41 OECD and the EU countries are ranked according to their environmental performance on a scale from 1 to 10 (Stiftung 2020). Higher index scores indicate better environmental performance, based on the criteria of the report.  For example, 1 is for the countries in which “Environmental concerns have been largely abandoned,” and 10 is given to countries whose “Environmental policy goals are ambitious and effectively implemented as well as monitored within and across most relevant policy sectors that account for the largest share of resource use and emissions” (Stiftung 2020).  In this article, SGI scores are utilized in two sections.  First, it is used in section E where we analyze the starting periods of the related decreasing trends of the CO2 emissions per capita.  Furthermore, it is also used in section F where we compare and explain the carbon emission per capita levels of the countries in 2014.

\newpage
\chapter{ANALYSIS OF THE SELECTED COUNTRIES}
\vspace*{0.4 cm}
\doublespacing
\indent In this section, we analyze the carbon emission per capita emission trends of twenty-four selected countries to assess whether the Environmental Kuznets Curve (“EKC”) theory is applicable.  In order to assess the validity of the EKC in these countries, we plot the carbon emission per capita level of each country together with its income level, measured by GDP per capita, for the time-period between 1960-2016.  
Out of these twenty-four countries, fifteen of them are high-income countries and the other nine countries are referred to as non-high income countries in this article. The related GDP per capita of the selected high-income countries are all over 20,000 USD. The fifteen selected high-income countries that are analyzed in section (i) below in this article are Switzerland, Denmark, Australia, USA, Canada, Austria, Japan, Finland, Germany, Belgium, France, UK, Italy, Spain and Portugal, respectively. 
On the other hand, the related GDP per capita of the nine selected non-high income countries are all below 12,000 USD.  The nine selected non high-income countries that are analyzed in section (ii) below in this article are Brazil, Malaysia, South Africa, Colombia, China, Peru, Egypt, India and Pakistan, respectively, 
\\
i.	High income countries :\\
\indent The graphs showing the carbon emission per capita vs GDP per capita trends for the fifteen high-income countries are presented in Figures 1-15.  The time-period for these graphics are mostly from 1990 to 2016 unless the information is not available for a specific country for this period in the World Bank’s database.  The information is unavailable only for a few of these selected countries, either at the beginning or at the end of the time-period.  The graphs for the high-income countries are ordered by the GDP per capita level of each country in 2016 in descending order.
When all the graphs for the selected high-income countries are analyzed, clearly, in all of these high income countries carbon emission per capita is increasing together with the increase in GDP per capita until a specific high level of GDP per capita is reached in the related country. When the specific high level of GDP per capita is reached, then, the carbon emission per capita levels off and starts to decline even though the GDP per capita continues to increase, forming an inverted U-shaped curve as defined by the EKC theory.  
For each one of the selected fifteen high-income countries, we present a brief analysis below:\\
1)	Switzerland:\\
\indent In 2016, Switzerland had the highest GDP per capita (77,026 USD) among the selected countries whereas its carbon emission per capita emission level was rather low at 4.12 metric tons per capita.  

 When we analyze Switzerland’s carbon emission per capita vs GDP per capita graph (Figure 1), we see that carbon emission per capita reached its peak as early as in 1973 at 7.33 metric tons per capita when the GDP per capita was high at 53,777 USD.  Carbon emission per capita started to decline after this peak point at 7.33 metric tons per capita in 1973 until it dropped down to 4.12 metric tons per capita in 2016.

The inverted U-shape curve for Switzerland in Figure 1 validates the EKC theory as its carbon emission per capita is at an increasing trend as GDP per capita was increasing and then it turned into a declining trend after reaching its peak point when the GDP per capita reached 53,777 USD in 1973. \\
2)	Denmark:\\
\indent Denmark had the second highest GDP per capita after Switzerland among the selected countries in 2016 at 61,878 USD. Even though Denmark’s carbon emission per capita emission level was higher than Switzerland’s in 2016, it was still rather low when compared to the emissions of the other selected countries at 5.55 metric tons per capita.  
In Denmark, carbon emission per capita emission reached its peak in 1996 at 13.71 metric tons per capita when the GDP per capita was high at 50,262 USD.  The trend of carbon emission per capita started to decline after this peak point and it reached 5.55 metric tons per capita in 2016.
In Figure 2, Denmark’s carbon emission per capita graph clearly has an inverted U- shape that validates the EKC theory.\\
3)	Australia:\\
\indent Australia had the third highest GDP per capita after Switzerland and Denmark among the selected countries in 2016 at 55,729 USD. However, carbon emission per capita level of Australia in 2016 (15.54 metric tons per capita) is substantially higher when compared to the levels of Switzerland (4.12 metric tons per capita) and Denmark (5.55 metric tons per capita). 
Australia’s EKC curve presented an increasing trend of carbon emission per capita as the GDP per capita increased from 1960 to 2010.  In 2010, carbon emission per capita reached its peak point at 17.74 metric tons per capita where the GDP per capita level was 52,022 USD. From 2011 onwards, even though the GDP per capita of the country continued to increase, carbon emission per capita started its declining trend where it fell to 15.54 metric tons per capita level in 2016 when the GDP per capita was at 55,729 USD.  
It is important to note that, the trend of Australia’s carbon emission per capita trend is in line with those of the Switzerland and Denmark in the context that carbon emission per capita starts to drop when the country reaches its country specific high GDP per capita level, which is over 50,000 USD.  However, it is also important to note that overall carbon emission per capita level of Australia (15.54 metric tons per capita) is relatively much higher than Switzerland’s (4.12 metric tons per capita) and Denmark’s (5.55 metric tons per capita).  Despite the significant difference in the overall carbon emission per capita levels with these countries, Australia still went through a similar pattern of carbon emission per capita with Switzerland and Denmark where the CO2 per capita emission started to decline when GDP per capita reached over 50,000 USD. 
Australia’s carbon emission per capita emission curve in Figure 3 clearly presents the typical characteristics of the EKC theory with an inverted U-shaped curve.\\
4)	USA:\\
\indent USA had the fourth highest GDP per capita among the selected countries in 2016 at 52,556 USD. However, USA’s carbon emission per capita level in 2016 is very high (15.50 metric tons per capita), similar to the level of Australia. 
USA’s carbon emission per capita vs GDP per capita graph is presented in Figure 4.  Carbon emission per capita reached its peak point at 22.5 metric tons per capita in 1973 when GDP per capita was at 25,794 USD.  However, the shape of the curve was volatile afterwards and carbon emission per capita started its consistent decreasing trend only after 2001 when GDP per capita reached its specific high enough level at 44,729 USD.  In 2000, carbon emission per capita was at 20.2 metric tons per capita before it started its declining trend in 2001. However, despite the declining curve trend, carbon emission per capita of the USA was still relatively high when compared with other countries even at its lowest point (15.50 metric tons per capita) in 2016 when GDP per capita was at 52,556 USD.
As shown in Figure 4, USA’s graph also has the typical characteristics of the EKC. The shape of the curve is an inverted U-shape and it validates the EKC theory.\\
5)	Canada:\\
\indent Canada had the fifth highest GDP per capita among the selected countries in 2016 at 50,193 USD and, furthermore, its carbon emission per capita was rather high like Australia and USA at 15.09 metric tons per capita.
The carbon emission per capita trend for Canada (Figure 5) has two main similarities with the USA’s graph.  First similarity is the long time lag between the peak level of carbon emission per capita and the start of the consistent decreasing trend afterwards.  Carbon emission per capita reached its peak point in 1980 at 18.08 metric tons per capita when GDP per capita was at 29,357 USD.  However, the consistent decreasing trend started only after twenty-seven years in 2007 when the carbon emission per capita level was at 17.39 metric tons per capita and the GDP per capita was at 48,534 USD. The second similarity with the USA’s carbon emission trend is their relatively high carbon emission per capita levels in 2016.  In fact, Canada’s lowest carbon emission per capita was in 2016 at 15.09 metric tons per capita level, similar to the level of USA, which was also high at 15.50 metric tons per capita in the same year.
The shape of Canada’s carbon emission per capita trend is slightly different from the other high-income countries.  Although Canada’s carbon emission per capita curve shape is also an inverted U-shape similar to the other high-income countries’ curves, its inverted U-shape is slightly different from the rest.  This difference in shape is a result of Canada’s dependence on petroleum, natural gas, and hydroelectricity for its economic development. Canada’s economy is relatively more energy intensive when compared to the other high- income countries.  It ranks fourth among the top energy producers of petroleum and total liquids in the world, behind only the United States, Saudi Arabia, and Russia and, furthermore, it exports a significant amount of the produced energy to the United States (US Energy Information Administration 2020). 
As shown in Figure 5, Canada’s carbon emission per capita graph also has an inverted U-shape and it validates the EKC theory. Since Canada’s economy largely depends on energy production, we see a slightly different inverted U-shaped curve in its graph.\\
6)	Austria:\\
\indent Austria has one of the highest GDP per capita levels within the selected European countries in this article.  In 2016, the GDP per capita of Austria was at 48,260 USD and the carbon emission per capita was at 7.03 metric tons per capita.
In Austria, carbon emission per capita increased together with the increase in GDP per capita until it reached its peak year in 2005 at 8.97 metric tons per capita when the GDP per capita was 44,638 USD.  The carbon emission per capita started to decline after this peak point. In 2016, carbon emission per capita level dropped to 7.03 metric tons per capita when the GDP per capita level was at 48,260 USD.  In 2016, Austria’s carbon emission per capita level was much less than Australia’s, USA’s and Canada’s but higher than Switzerland’s and Denmark’s.
As shown in Figure 6, Austria’s graph also has the typical characteristics of the EKC. The shape of the curve is an inverted U-shape and it validates the EKC theory.\\
7)	Japan:\\
\indent Japan has the highest GDP per capita level within the selected Asian countries and it ranks as the seventh highest in income level within all the selected countries in this article.  In 2014, the GDP per capita of Japan was at 47,403 USD and the carbon emission per capita was at 8.94 metric tons per capita.
At first, Japan’s carbon emission per capita also increased with the increase in its GDP per capita, as suggested by the EKC hypothesis. However, when it reached its peak levels, it stayed at these levels for a very long time (forty years) as shown in Figure 7.  Japan’s carbon emission per capita level reached 8.47 metric tons per capita in 1973 and stayed at these high levels for forty years until it reached its peak point in 2004 at 9.9 metric tons per capita when GDP per capita was at 43,672 USD.  After reaching this peak point in 2004, carbon emission per capita started to decrease in line with the relative decrease in GDP per capita. In 2013, GDP per capita reached a higher level when compared with the past 10 years at 46,249 USD and, accordingly, carbon emission per capita increased to 9.76 metric tons per capita.  In 2014, however, a consistent declining trend of carbon emission per capita started even though GDP per capita (46,484 USD) was increasing.  Carbon emission per capita dropped to 8.9 metric tons per capita level where GDP per capita was at 47,403 USD in 2016.  
Japan’s carbon emission per capita curve (Figure 7) shape also supports the EKC theory because the carbon emission per capita increased together with the increase in GDP per capita until the country specific high level of GDP per capita was reached in 2013 and, then, the carbon emission per capita started to decline from 2014 onwards.  However, Japan’s EKC trend is slightly different from the majority of the high-income countries’ curves because of the long time lag between the year when the peak level of carbon emission per capita was reached (2004) and the year when the decreasing trend of carbon emission per capita started (2014).  One of the reasons for this long time lag between the peak year and the year of the start of the declining trend is the GDP per capita trend within this period.  The GDP per capita also dropped and went up again within this period, affecting the inverted U-shape of the carbon emission per capita curve, accordingly.
In fact, there is a similarity between Japan’s and Canada’s carbon emission per capita curves. In both of these countries, carbon emission per capita levels stayed at high levels for a long time before starting to decline. The carbon emission level per capita level stayed high in Canada for twenty-seven years whereas this period lasted forty years in Japan. 
There are two main reasons for Japan’s very long, sticky period at high levels of carbon emission per capita.  The first reason is Japan’s high energy usage. According to 2019 statistics of the US Energy Information Administration, Japan is a major energy importer in the world as the fifth-largest oil consumer.  Furthermore, it is the largest liquefied natural gas (LNG) importer, the fourth-largest crude oil importer and the third-largest coal importer behind China and India. Japan historically used to have higher shares of energy production in nuclear energy until the Fukushima nuclear energy accident in 2011. After the nuclear accident, energy fuel mix has shifted to natural gas, oil and renewable energy. Although the share of oil in energy production declined from about 80\% in the 1970’s to 40\% in 2019 as a result of declining and aging population, high energy efficiency measures, and an expanding fleet of hybrid and electric vehicles, it is still the major source of energy production.  Coal still has 26\% share in total energy consumption whereas the increasing share of natural gas has reached 21\% of total primary consumption.  Before the 2011 earthquake, Japan was the third-largest nuclear power producer after USA and France. Almost 13\% of Japan’s total energy production was produced from nuclear energy in 2010 whereas this share decreased sharply to 3\% in 2010 after the earthquake.  The share of nuclear energy will be increasing again in the near future as per the government’s plan to reduce energy imports.  Renewable energy is about 10\% of Japan’s energy consumption and its share is expected to increase (US Energy Information Administration 2020).
The second reason for Japan’s long period of high CO2 per capita emission level is the “lost decade” which started with the economic stagnation caused by the burst of the asset price bubble in early 1990’s (Callen and Ostry 2013).  This economic stagnation slowed down the GDP per capita growth of Japan when compared with other countries and resulted in lengthening the period where the CO2 per capita emission stayed high.  Despite this long period of high carbon emission per capita, the shape of the CO2 per capita curve of Japan still validates the EKC theory.\\
8)	Finland:\\
\indent Finland is also one of the selected high-income European countries in this article.  In 2016, the GDP per capita of Finland was at 46,750 USD and the carbon emission per capita was at 8.35 metric tons per capita.
Finland’s carbon emission per capita curve trend is shown in Figure 8.  In Finland’s curve, at the beginning of the related time-period carbon emission per capita increased together with the GDP per capita, as it is also stated in the EKC theory.  In 1980, carbon emission per capita reached its high level of 12.19 metric tons per capita level when GDP per capita was at 25,512 USD.  Carbon emission per capita stayed around this level until it reached its peak level at 13.17 metric tons per capita when GDP per capita was at 42,707 USD in 2003.  In line with the EKC theory, the decreasing trend started after this peak point was reached in 2003.  In 2016, carbon emission per capita level dropped to 8.35 metric tons per capita level when GDP per capita was at 46,750 USD.
Finland’s carbon emission per capita emission curve shape presented in Figure 8 is an inverted U-shape and it clearly validates the EKC theory.\\
9)	Germany:\\
\indent In 2016, the GDP per capita of Germany was at 45,960 USD and the carbon emission per capita was at 8.84 metric tons per capita.
In Figure 9, Germany’s carbon emission per capita vs GDP per capita graph is presented.  The carbon emission curve is graphed between the years of 1991 to 2016 because Germany’s carbon emission per capita information is available only after 1991 due to its political history. In 1945, after the Second World War, the country was divided into East Germany and West Germany and German reunification did not take place until 1990. As a result, the carbon emission per capita data is available only from 1991 onwards after the reunification.  
In 1991, the reunified Germany’s GDP per capita was at 33,836 USD and its CO2 per capita was at 11.62 metric tons per capita.  Germany’s carbon emission per capita curve presented a declining trend since 1991 since its GDP per capita was already high enough in 1991 for the start of the declining trend.  Germany’s CO2 per capita dropped to 8.84 metric tons per capita in 2016 from 11.62 in 1991.  
Germany’s carbon emission curve also supports the EKC theory, which states that when a high-enough GDP per capita is reached, a declining trend of CO2 per capita is observed even though GDP per capita continues to increase.\\
10)	Belgium:\\
\indent In 2016, the GDP per capita of Belgium was at 45,943 USD and the carbon emission per capita was at 8.55 metric tons per capita.
In Belgium’s carbon emission per capita graph (Figure 10), carbon emission per capita had increased from 1960 to 1973.  In 1973, carbon emission per capita reached its peak level at 14.26 metric tons per capita when the GDP per capita of Belgium was at 22,868 USD.  In 1979, when GDP per capita reached 25,946 USD, carbon emission per capita was very close to its peak level again at 14.24 metric tons per capita and, after this year, from 1980 onwards, carbon emission per capita followed a declining trend.  In 2016, carbon emission per capita dropped to 8.55 metric tons per capita whereas GDP per capita increased to 45,943 USD.
Belgium’s CO2 per capita vs GDP per capita graph clearly has the characteristics stated in the EKC theory where carbon emission per capita had increased together with the GDP per capita until the peak level was reached and, then, a declining trend of CO2 per capita was observed despite the increase in GDP per capita. \\
11)	France:\\
\indent In 2014, the GDP per capita of France was at 41,481 USD and the carbon emission per capita was at 4.57 metric tons per capita.  Since the carbon emission per capita data in the World Bank’s database for France and Italy is only up to 2014, France’s graph in Figure 11 presents the carbon emission per capita curve from 1960 to the last available data in 2014.
In 1960, the carbon emission per capita level in France was at 5.82 metric tons per capita whereas its GDP per capita was at 12,744 USD.  However, in 1973, carbon emission per capita reached its peak level at 9.71 metric tons per capita when the GDP per capita was at 22,843 USD.  The declining trend started after 1979 when the carbon emission per capita reached its peak level at 9.6 metric tons per capita and the GDP per capita was at 26,578 USD.  After a sharp declining trend, in 2014 the CO2 per capita dropped to 4.57 metric tons per capita despite the increase in GDP per capita to 41,481 USD.
Similar to the graphs of the other European countries that are analyzed above in this article, France’s carbon emission per capita vs GDP per capita graph in Figure 11 also presents an inverted U-shaped curve and supports the hypothesis of the EKC theory.\\
12)	United Kingdom (“UK”):\\
\indent In 2016, the GDP per capita of the United Kingdom was at 42,500 USD and the carbon emission per capita was at 5.78 metric tons per capita.
The carbon emission per capita vs GDP per capita curve for the United Kingdom is presented in Figure 12. Carbon emission per capita in the United Kingdom slightly increased from the 11.15 metric tons per capita level in 1960 (when the GDP per capita was at 13,934 USD) to 11.82 metric tons per capita at its peak level in 1971 as the GDP per capita increased to 18,474 USD. However, starting from 1974 onwards, carbon emission per capita in the United Kingdom started to decline and, in 2016, it dropped to 5.78 metric tons per capita, approximately half of its peak level in 1971.  
In summary, UK’s carbon emission per capita curve also clearly supports the hypothesis of the EKC theory which states that carbon emission per capita starts to decline after reaching its peak level at a certain high GDP per capita level.\\
13)	Italy:\\
\indent Similar to France, the carbon emission per capita data in the World Bank’s database for Italy is available also only up to 2014.  In 2014, the GDP per capita of Italy was at 33,667 USD and the carbon emission per capita was at 5.27 metric tons per capita.
In Italy, carbon emission per capita was at 2.18 metric tons per capita in 1960 and the GDP per capita was at 10,879 USD.  Carbon emission per capita reached its peak point (8.22 metric tons per capita) when GDP per capita more than tripled from its level in 1960 to 37,227 USD in 2004. As shown in Figure 13, from 2005 onwards, carbon emission per capita started to decline and in 2014 it dropped down to the 5.27 metric tons per capita level.  
Italy’s carbon emission per capita vs GDP per capita graph also clearly presents an inverted U-shaped curve as shown in Figure 13.  As a result, the trend in Italy also supports the hypothesis of the EKC theory.\\
14)	Spain:\\
\indent In 2016, the GDP per capita of Spain was at 31,449 USD and the carbon emission per capita was at 5.25 metric tons per capita.
In 1960, the carbon emission per capita in Spain was low at 1.61 metric tons per capita and GDP per capita was at 7,376 USD.  In the coming years, the carbon emission per capita increased together with the increase in the GDP per capita until it reached its peak level in 2005.  In 2005, GDP per capita was at 31,029 USD and carbon emission per capita was at its peak at 8.10 metric tons per capita.  From 2006 onwards, carbon emission per capita trend showed a consistent declining trend and it dropped down to 5.25 metric tons per capita level in 2016.
As shown in Figure 14, Spain’s carbon emission per capita vs GDP per capita graph also presents an inverted U-shaped curve, similar to the other high-income countries’ graphs that are analyzed above in this article.  Therefore, Spain’s carbon emission per capita curve also has the related characteristics of a typical EKC and supports the hypothesis of this theory. \\
15)	Portugal:\\
\indent In 2016, the GDP per capita of Spain was at 22,534 USD and the carbon emission per capita was at 5.25 metric tons per capita.
In 1960, Portugal’s GDP per capita was at 4,501 USD and carbon emission per capita was at 0.93 metric tons per capita.  Portugal’s carbon emission per capita level increased together with the increase in its GDP per capita until it reached its peak point in 1999 at 6.31 metric tons per capita, approximately 6.8 times its level in 1960. When the carbon emission per capita level was at its peak in 1960, the GDP per capita was at 20,853 USD level, which was approximately 4.6 times its level in 1960.  Carbon emission per capita stayed around this peak level for 6 years until 2005 and it started to decline only after 2006 onwards.  In 2016, Portugal’s GDP per capita increased to 22,534 USD whereas its carbon emission per capita dropped to 4.72 metric tons per capita.
As shown in Figure 15, Portugal’s carbon emission per capita trend also presents the characteristic inverted U-shaped curve of the EKC hypothesis, similar to all the other high-income countries that were analyzed in this article.\\
16)	Conclusion for the analysis on high-income countries:\\
\indent As the analyses of the long-term carbon emission per capita trend for all the selected fifteen countries consistently showed, the declining trend of the carbon emission curve starts only after the related, country specific high GDP per capita level is reached in each of these countries. 
Furthermore, we observe that the specific high level of GDP per capita year differs from country to country.  Therefore, we conclude that the environmental protection policies issued by the international agencies are ineffective in initiating a decrease in these high-income countries’ carbon emission levels.  If the policies of the international institutions were successful in decreasing the carbon emission levels, then, the carbon emission per capita levels in these countries would have started to drop around the same year, in coherence with the timing of the effectiveness of these policies.  However, when we analyze the carbon emission per capita emission trends of these selected fifteen high-income countries, it is clear that there is not a correlation between the decreasing trend of the carbon emission and the timing of these international institutions policies. The CO2 per capita vs GDP per capita graphs clearly show that the carbon emission starts to decline in different years in these selected high-income countries. 
The results of the analyses of these fifteen selected high-income countries show clearly that the major determinant for the initiation of the declining carbon emission trend is reaching the related specific high-income level per capita relevant for each country.  All the graphs of these selected high-income countries are in accordance with the inverted U-shaped curve shape hypothesis of the EKC theory.  Also, the results of the analyses in this section also indicates that, unfortunately, international policies are ineffective in ensuring a decrease in the carbon emission levels of these high-income countries.

\pagebreak

\begin{figure}[H]
	\includegraphics[scale=0.8]{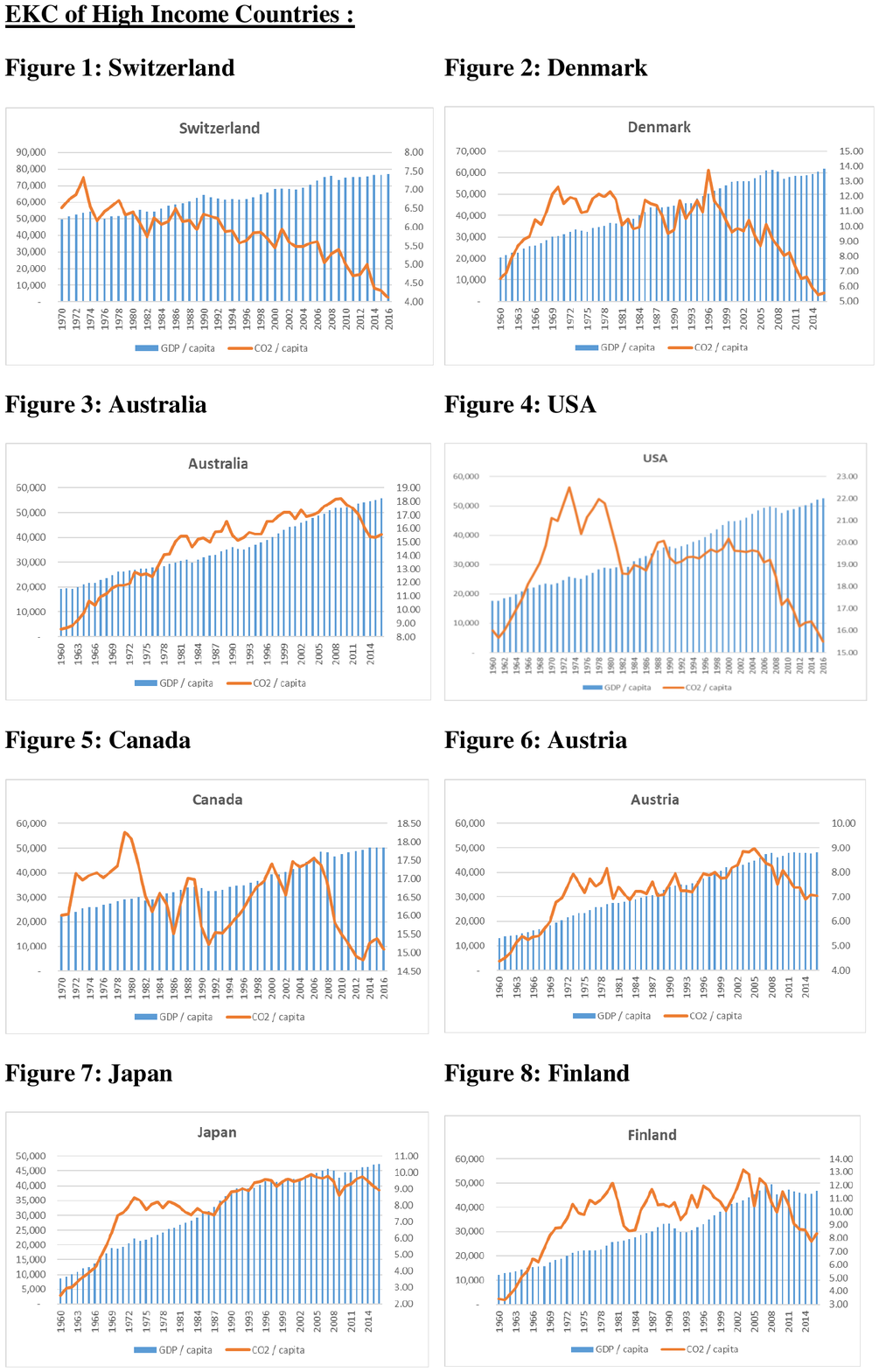}
\end{figure}

\pagebreak
\begin{figure}[H]
	\centering
	\includegraphics[scale=0.8]{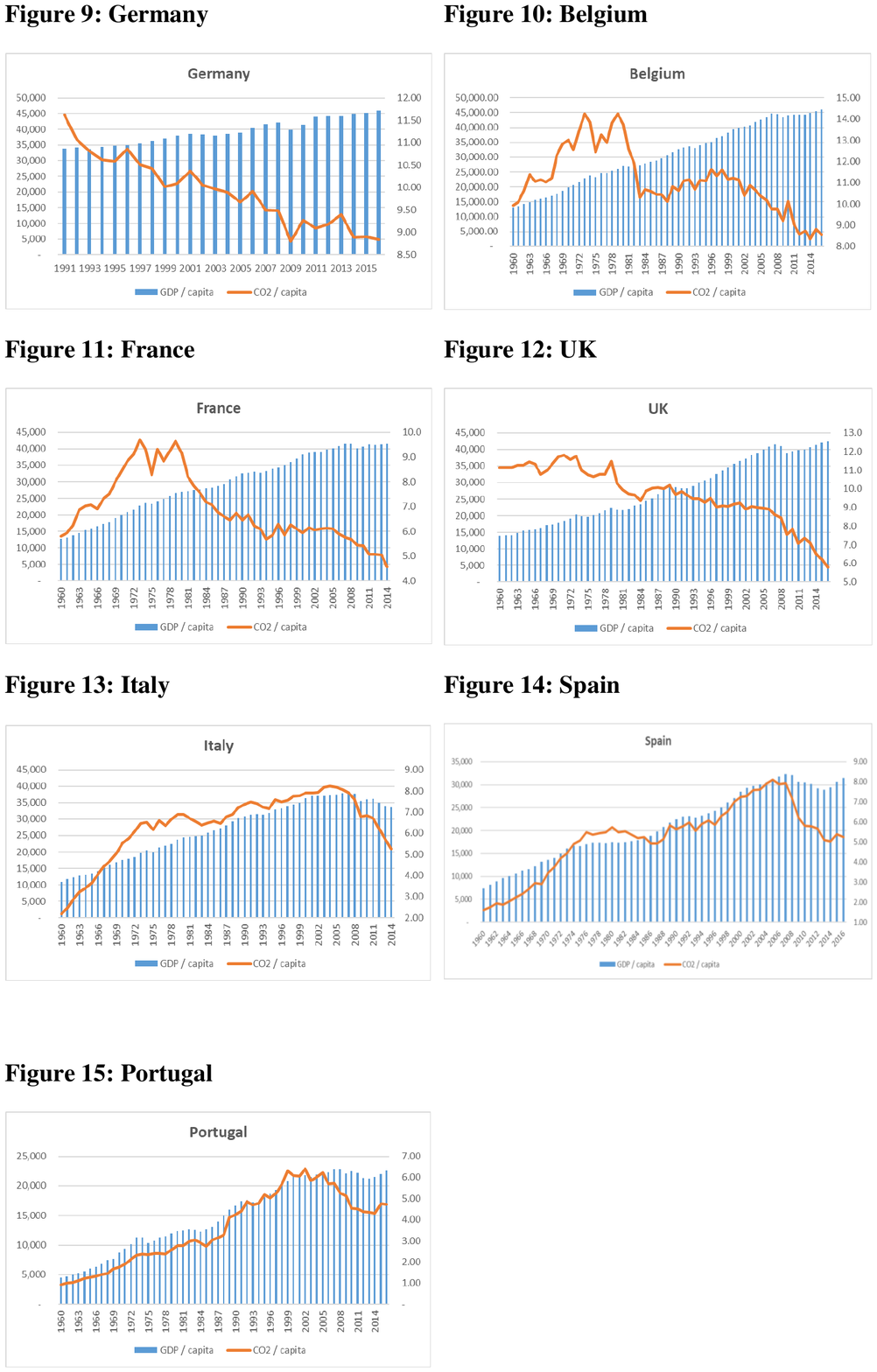}
\end{figure}

\pagebreak

ii.	Non - high income countries : \\
\indent Carbon emission per capita vs GDP per capita graph of the nine non-high income countries are analyzed in this section.  The nine non-high income countries are: Brazil, Malaysia, South Africa, Colombia, Peru, China, Egypt, India and Pakistan. The carbon emission curves and the GDP per capita trends of these countries are presented in Figures 16-24.
In general, the GDP per capita of these selected non-high income countries are relatively very low when compared to the high-income countries.  In fact, the average GDP per capita for these nine countries in 2016 was only 7,076 USD.  As a result, most of these countries still have an increasing trend of carbon emission because their economies are still growing and their GDP per capita is still increasing, accordingly.  The EKC theory states that carbon emission per capita will start to decline only when the GDP per capita reaches a certain country-specific high level.  Since the GDP per capita levels of these non-high income countries are still in the increasing trend stage, carbon emission per capita levels are also still at the increase stage and the peak points are not reached yet\\
\indent The graphs of these nine non-high income countries are analyzed below:
\\
1)	Brazil:\\
\indent In 2016, the GDP per capita of Brazil was at 10,966 USD and the carbon emission per capita was at 2.24 metric tons per capita.
Although Brazil has the highest GDP per capita among these selected nine low-income countries, carbon emission per capita is still in the steady increase trend since the level of GDP per capita has not reached its country specific high level yet.  For example, the GDP per capita for Portugal, the country with the lowest GDP per capita among the high-income countries, is 22,534 USD in 2016, which is approximately the double of Brazil’s GDP per capita.
The carbon emission level for Brazil increases from its 0.65 metric tons per capita level in 1960 (when GDP per capita was 3,417 USD), to 2.24 metric tons per capita level in 2016 when GDP per capita was at  10,966 USD.  
Brazil’s CO2 per capita vs GDP per capita graph is presented in Figure 16. The carbon emission per capita is still in the increasing phase, in line with its increasing GDP per capita trend.  Therefore, for Brazil, it is too early at this stage to state whether its carbon emission curve will be in line with the hypothesis of the EKC theory.  However, the increasing trend phase of the curve is in line with the first part of the EKC theory, which states that the carbon emission per capita is expected to increase until it reaches a country-specific high-income per capita level. \\
2)	Malaysia:\\
\indent In 2016, the GDP per capita of Brazil was at 11.244 USD and the CO2 per capita emission was at 8.09 metric tons per capita.
Malaysia’s carbon emission per capita curve in Figure 17 presents a strong increasing trend for CO2 per capita as its GDP per capita is increasing.  It would be fair to say that, it is too early at this GDP per capita level to state whether Malaysia’a carbon emission per capita trend will support the hypothesis of the EKC theory because the GDP per capita is still at a relatively low level (11,244 USD) in 2016. However, as it is in the case for Brazil, the increasing trend phase of the curve is in line with the first part of the EKC theory, which states that the carbon emission per capita is expected to increase until it reaches a country-specific high-income per capita level.\\
3)	South Africa:\\
\indent The income per capita level of South Africa is still at a rather low level in 2016, at 7,477 USD.  In fact, the GDP per capita level of South Africa is at almost one-third of the income per capita level of Portugal (22,534 USD), which has the lowest GDP per capita among the selected high-income countries. South Africa’s carbon emission per capita level is at 8.48 metric tons per capita.
South Africa is the largest coal producer in its region.  It has the tenth-largest recoverable coal reserves in the world and 75\% of the total coal reserves in Africa.  Energy consumption is increasing in South Africa as the economy is growing and the government is building more coal-fired power stations to meet the demand. However, similar to Canada, South Africa is also an energy exporter.  It is the fifth-largest coal exporter in the world and exports 30\% of its coal production to countries in Asia, especially to India.  
In brief, South Africa is a developing country where GDP per capita is still rather low and the demand for energy is increasing due to its growing economy.  Furthermore, South Africa is also exporting a significant amount of produced energy to other countries and its economic growth is dependent on these exports.  
As a result, carbon emission per capita in South Africa is still in an increasing trend as shown in Figure 18. It is still too early at this stage to state whether the carbon emission per capita curve of South Africa will be in line with the EKC hypothesis because its GDP per capita level is still at rather low levels and it is still a developing country. However, the increasing trend phase of the curve is in line with the first part of the EKC theory, which states that the carbon emission per capita is expected to increase until it reaches a country-specific high-income per capita level. \\
4)	Colombia:\\
\indent Similar to South Africa, the income per capita level of Colombia is still rather low at 7,634 USD in 2016.  However, Colombia’s carbon emission per capita level is at a much lower level, (2.03 metric tons per capita) when compared to the one of South Africa (8.48 metric tons per capita). 
According to 2019 statistics of the US Energy Information Administration, Colombia is South America’s largest coal producer and third-largest oil producer. However, even though it is such a major coal and oil producer, Colombia uses hydropower for most of its domestic electricity needs and exports coal and oil mainly to USA.  Therefore, its carbon emission per capita is much lower when compared to South Africa’s carbon emission.
As shown in Figure 19, Colombia’s carbon emission per capita is increasing as its GDP per capita is increasing.  However, as it was also the case in the other non-high income countries that are analyzed above, it is still too early at this stage to state whether the carbon emission per capita curve of Colombia will be in line with the EKC hypothesis because its GDP per capita level is still at rather low levels. However, the increasing trend phase of the curve is in line with the first part of the EKC theory, which states that the carbon emission per capita is expected to increase until it reaches a country-specific high-income per capita level.\\
5)	China:\\
\indent In Figure 20, China’s carbon emission per capita trend is presented together with its GDP per capita trend. In 2016, GDP per capita of China is still rather low at 6,908 USD and carbon emission per capita level is at 7.18 metric tons per capita level.
In China’s graph, carbon emission per capita trend is increasing as its GDP per capita trend is increasing.  In 1960, carbon emission per capita was at 1.17 and GDP per capita was at only 192 USD.  In 2016, carbon emission per capita increased 6.1 times and reached 7.18, whereas GDP per capita increased almost 36 times to 6,908 USD.  As it is the case in the other non-high income countries analyzed above, it is still too early at this stage to state whether the carbon emission per capita curve of China will be in line with the EKC hypothesis because its GDP per capita level is still at rather low levels. However, the increasing trend phase of the curve is in line with the first part of the EKC theory, which states that the carbon emission per capita is expected to increase until it reaches a country specific high-income per capita level. \\
6)	Peru:\\
\indent In 2016, the income per capita level of Peru is still rather low at 6,262 USD and its carbon emission is at 1.86 metric tons per capita.
Peru is also similar to South Africa and Colombia in exporting energy sources despite its growing domestic needs.  Peru has rich hydrocarbons, oil, natural gas and coal reserves (US Energy Information Administration 2020).  
In 1960, Peru’s carbon emission was at 0.80 metric tons per capita and it increased 2.3 times to 1.86 metric tons per capita in 2016.  As shown in Figure 21, Peru’s carbon emission per capita is increasing as its GDP per capita is increasing.  However, as it was also the case in the other non-high income countries that are analyzed above, it is still too early at this stage to state whether the carbon emission per capita curve of Peru will be in line with the EKC hypothesis because its GDP per capita level is still at rather low levels. However, the increasing trend phase of the curve is in line with the first part of the EKC theory, which states that the carbon emission per capita is expected to increase until it reaches a country-specific high-income per capita level.\\
7)	Egypt:\\
\indent In 2016, GDP per capita of Egypt is still very low at 2,761 USD and carbon emission per capita level is at 2.53 level. In Figure 22, Egypt’s carbon emission per capita trend is presented together with its GDP per capita trend.  Egypt’s carbon emission curve is in an increasing trend as its GDP per capita is increasing.
In 1960, carbon emission per capita was at 0.60 and GDP per capita was at only 578 USD.  Although GDP per capita of Egypt is much higher than China’s (192 USD) in 1960, it is still rather low. In 2016, carbon emission per capita increased approximately 4.2 times and reached 2.53 whereas GDP per capita increased 4.8 times to 2,761 USD.  Despite the high increase rate in its GDP per capita, it is still too early at this stage to state whether the carbon emission per capita curve of Egypt will be in line with the EKC hypothesis because its GDP per capita level is still at rather low levels. However, the increasing trend phase of the curve is in line with the first part of the EKC theory, which states that the carbon emission per capita is expected to increase until it reaches a country specific high-income per capita level.\\
8)	India:\\
\indent In 2016, GDP per capita of India is rather low at 1,876 USD and CO2 per capita level is also low at 1.82 level. As it is presented in Figure 23, India’s CO2 per capita trend is increasing together with its GDP per capita. 
In 1960, CO2 per capita was at 0.27 and GDP per capita was at only 330 USD. In 2016, CO2 per capita increased approximately 6.7 times and reached 1.82 whereas GDP per capita increased 5.7 times to 1,876 USD.  
Despite the high increase rate in its GDP per capita, its level in 2016 is still low at 1,876 USD. Therefore, it is still too early at this stage to state whether the carbon emission per capita curve of India will be in line with the EKC hypothesis because its GDP per capita level is still at rather low levels.  However, the increasing trend phase of the curve is in line with the first part of the EKC theory, which states that the carbon emission per capita is expected to increase until it reaches a country specific high-income per capita level.\\
9)	Pakistan:\\
\indent In 2016, GDP per capita of Pakistan is rather low at 1,118 USD and CO2 per capita level is also low at 0.99 level. 
As it is presented in Figure 24, Pakistan’s CO2 per capita trend is also increasing together with its GDP per capita, similar to the rest of the selected non-high income countries. 
In 1960, CO2 per capita was at 0.31 and GDP per capita was at only 302 USD.  In fact, both the carbon emission per capita and the income per capita of Pakistan is very similar to that of India in 1960.  In 2016, CO2 per capita increased approximately 3.2 times and reached 0.99 whereas GDP per capita increased 3.7 times to 1,118 USD.  For the period of 1960-2016, carbon emission and income growth rates in Pakistan are high but relatively lower than India’s. 
Despite the substantial increase rate in its GDP per capita, its level in 2016 is still very low at 1,118 USD. As it was also the case in all of the other selected non-high income countries, it is still too early at this stage to state whether the carbon emission per capita curve of Pakistan will be in line with the EKC hypothesis because its GDP per capita level is still at rather low levels. However, the increasing trend phase of the curve is in line with the first part of the EKC theory, which states that the carbon emission per capita is expected to increase until it reaches a country specific high-income per capita level.\\
10)	Conclusion for the analysis on non-high income countries:\\
\indent Nine selected non-high income countries (Brazil, Malaysia, South Africa, Colombia, China, Peru, Egypt, India, Pakistan) are analyzed in the section above.  The carbon emission per capita of all these countries are still in an increasing trend because their GDP per capita levels are still at rather low levels.  In summary, it is still too early at this stage to state whether the carbon emission per capita curve of these selected non-high income countries will be in line with the EKC hypothesis due to their still rather low GDP per capita levels. However, the increasing trend phase of their carbon emission curves are in line with the first part of the EKC theory, which states that the carbon emission per capita is expected to increase until it reaches a country-specific high-income per capita level.

\pagebreak

\begin{figure}[H]
	\includegraphics[scale=0.8]{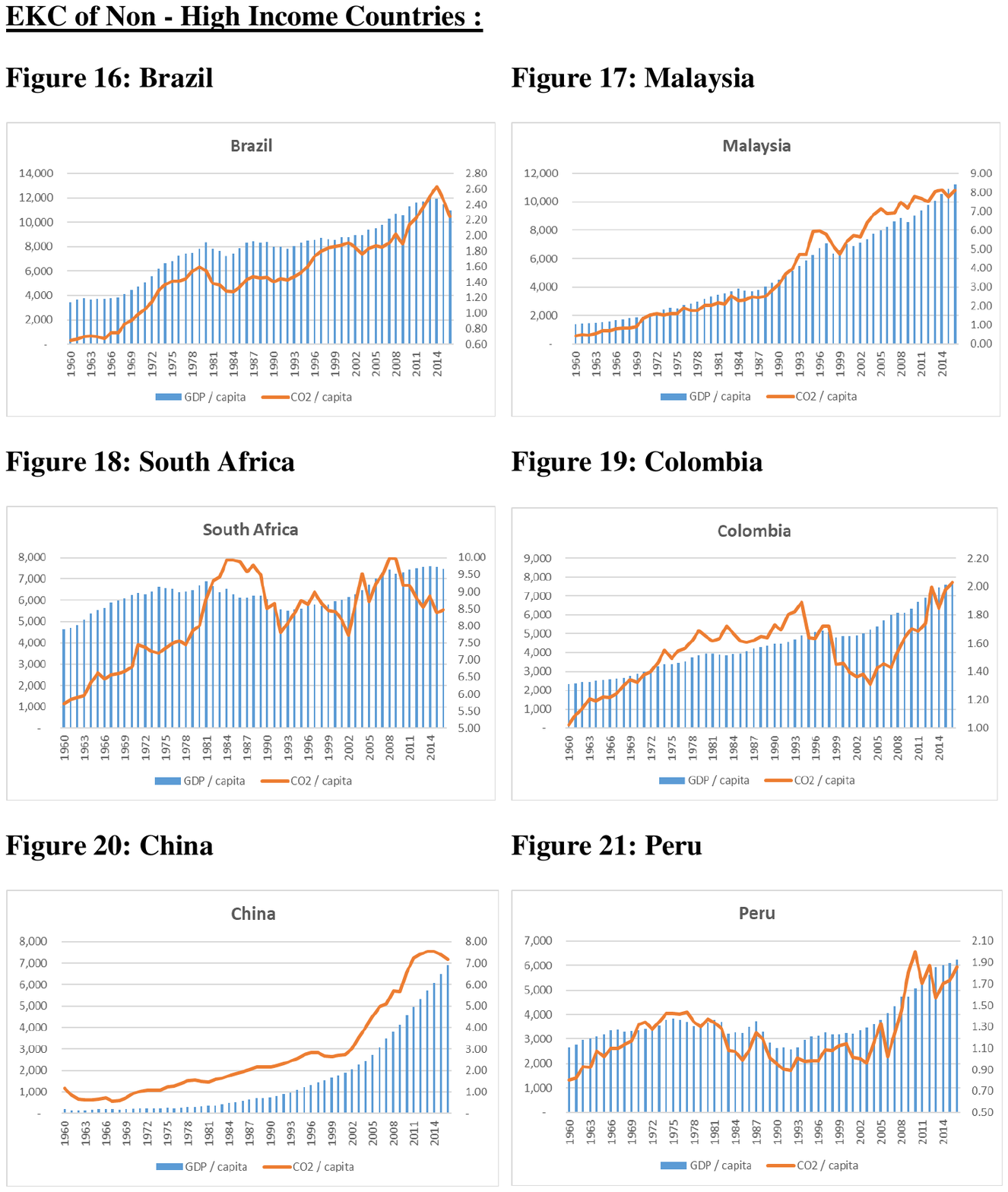}
\end{figure}
\pagebreak

\begin{figure}[H]
	\centering
	\includegraphics[scale=0.8]{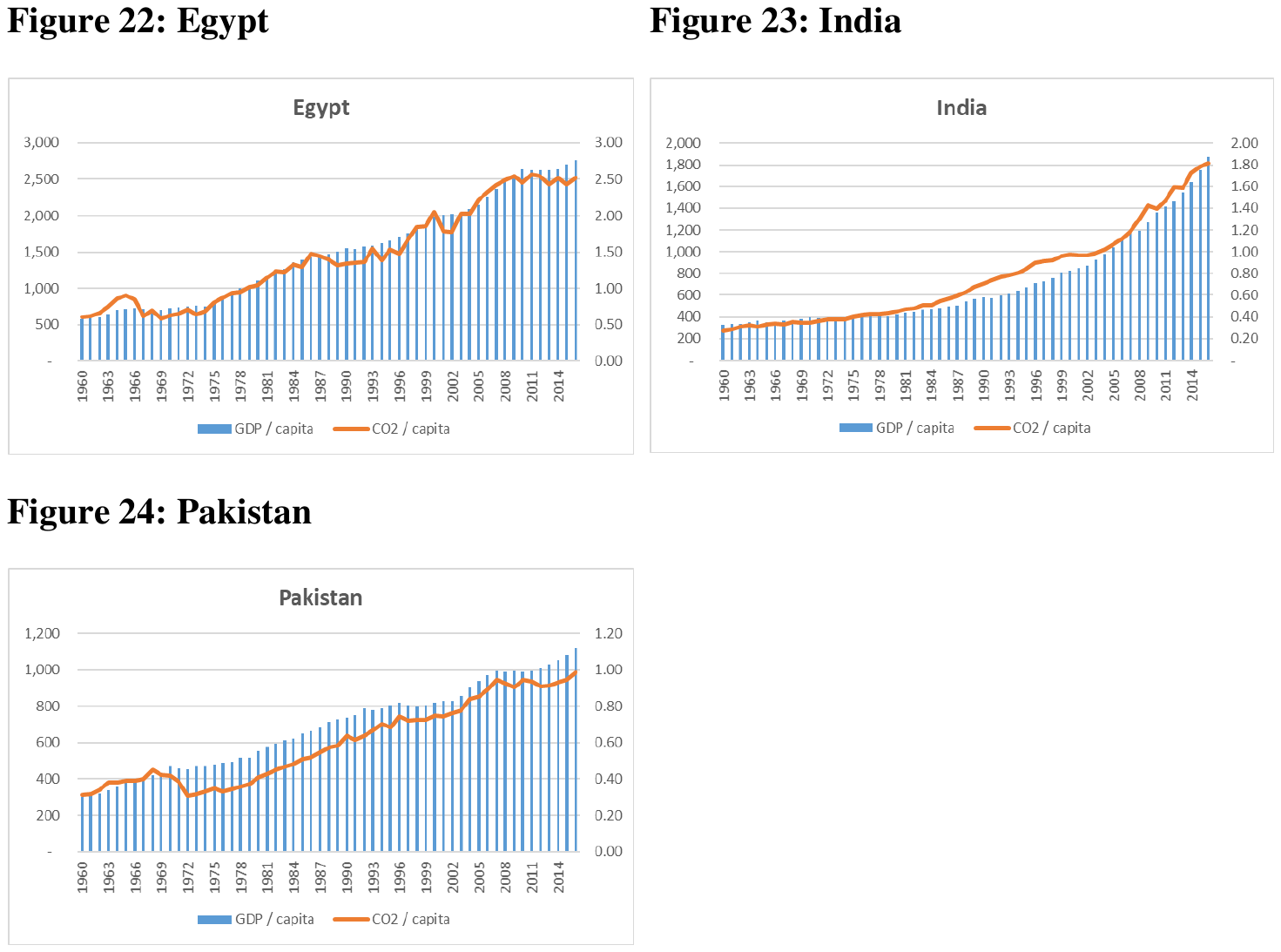}
\end{figure}
\pagebreak

\clearpage

\chapter{THE START OF THE CARBON EMISSION DECLINE}

\doublespacing
\indent In section D(i), the carbon emission trends of the fifteen selected high-income countries were analyzed and it was observed that for the high-income countries, carbon emission per capita level of each country started to decline only after the country specific high-income level, measured by GDP per capita, was reached. Furthermore, it was observed that the decline in the carbon emission levels started in different years and, as a result, it was inferred that the policies of the international institutions were ineffective in initiating the decline trends in carbon emission levels. In this section, the year of the start of the declining trend for the carbon emission per capita for these selected high-income countries is further analyzed.  \\
\indent In the table below, the year of the start of the decreasing trend, the related GDP of that year, the level of CO2 per capita and the SGI of 2018 for the high income countries (except for Germany that shows only a decreasing trend due to data unavailability) are summarized:

\begin{figure}[H]
	\centering
	\includegraphics[scale=0.6]{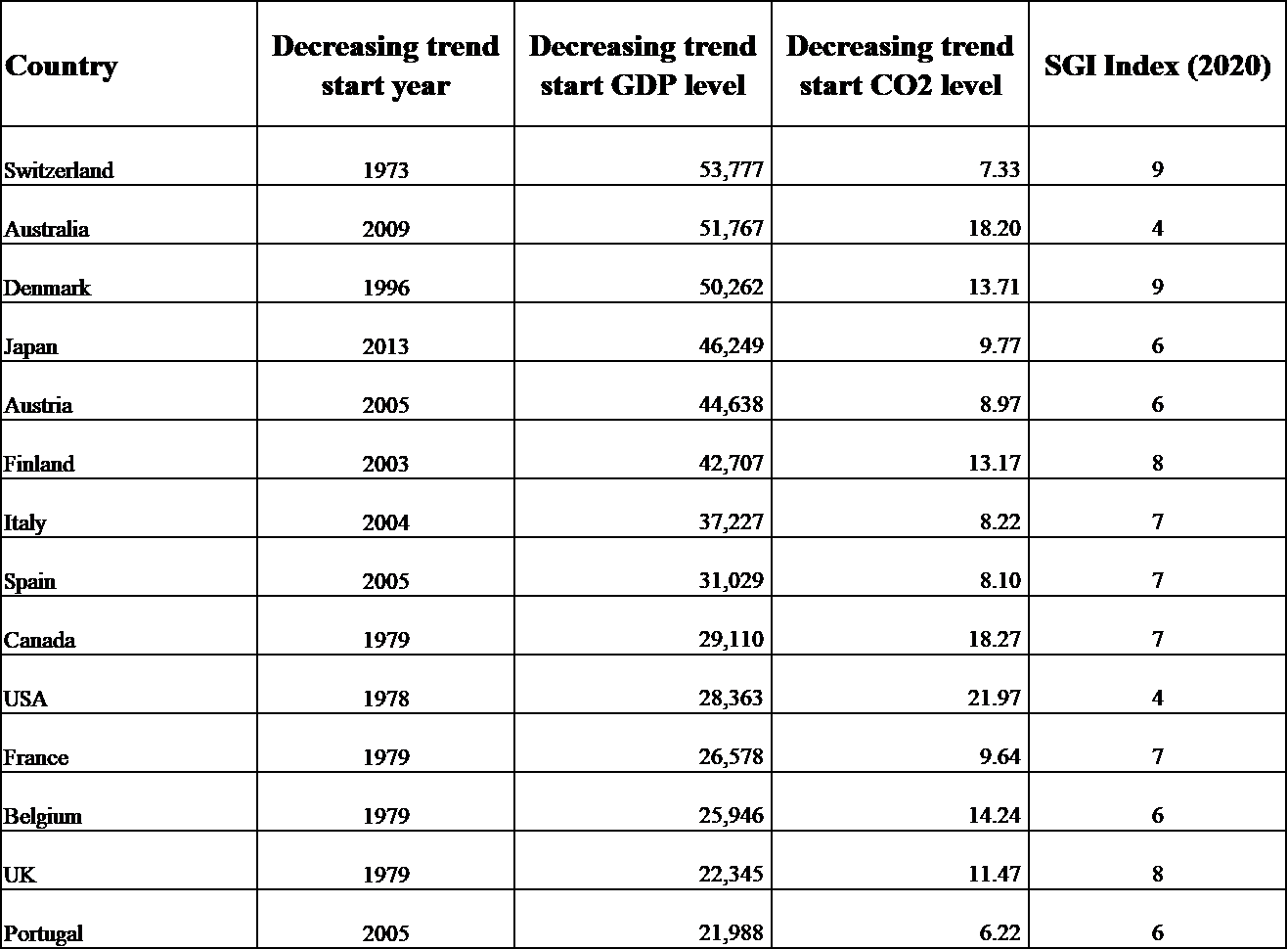}
\end{figure}
 
\indent The table above clearly shows that the year of the start of the declining trend is very different for the high-income countries. This difference of the year in which the carbon emission declining trend starts proves that the policies of the international institutions are unfortunately ineffective in initiating the start of the declining trend in these countries.  If the policies of the international policies were effective, then, the declining trend of the carbon emission curve would have started around the same years.  In the table above, we can see that the starting year of the decline in the carbon emission trend has a wide range from 1973 to 2013.  \\
\indent The country with the earliest carbon emission per capita decreasing trend is Switzerland.  Switzerland’s carbon emission per capita started to decrease after 1973 when its GDP per capita was at 53,777 USD. Furthermore, Switzerland is a very successful country in environmental protection with a SGI index score of 9. This awareness and responsiveness to carbon emission together with its high GDP per capita explains the reason for Switzerland’s earliest start of the decreasing trend in its carbon emission curve.\\
\indent On the other hand, Australia’s decreasing trend of carbon emission per capita started in 2009 when its GDP per capita was at 51,767 USD.  GDP per capita level of Australia was similar to that of Switzerland when its carbon emission per capita started to decrease.  However, SGI for Australia was quite low (4) when compared to the high score of Switzerland (9).  Furthermore, carbon emission per capita levels of these countries were also quite different at the time when their decreasing trends started.  Carbon emission per capita for Switzerland was at 7.33 metric tons per capita whereas that of Australia was very high at 18.20 metric tons per capita.  Therefore, it can be inferred that despite the differences in carbon emission sensitivity and performance of these countries, the carbon emission per capita started to decline only when the country specific high-income per capita was reached. \\
\indent Denmark and Finland were both European countries with very close levels of carbon emission per capita levels at the start of the decreasing trend of their carbon emission curves.  Denmark’s carbon emission level was at 13.71 metric tons per capita and Finland’s was at 13.17 metric tons per capita.  Both of these countries also have high SGI index scores where Denmark’s score is at 9 and Finland’s score is at 8.  However, despite these similarities in their carbon emission trends, sensitivities and performances, the starting year of their decreasing carbon emission trends were very different.  Denmark’s decreasing trend started relatively earlier in 1996 whereas Finland’s decline in the carbon emission per capita curve started 7 years later in 2003.  It is clear that the GDP per capita is the major differentiating factor for the time difference in the start of the declining carbon emission curves of these countries.  Denmark’s GDP per capita reached 50,262 USD in 1996 whereas Finland’s was at still 32,963 USD in that year.  In fact, carbon emission per capita for Finland started to drop only in 2003 only when its GDP per capita reached 42,707 USD which was its country specific high income level as per the hypotheses of EKC theory.\\
\indent The declining trends of the carbon emission curves of Italy and Spain start in 2004 and in 2005, respectively.  Their GDP per capita levels are both above 30,000 USD.  Italy’s GDP per capita is at 37,227 USD in 2004 and that of Spain is at 31,029 USD in 2005. Furthermore, their carbon emission per capita levels are similar, as well.  Carbon emission per capita for Italy is 8.22 metric tons per capita at the start of the decreasing trend in 2004 whereas that of Spain is at 8.10 metric tons per capita in 2005.  Their SGI index scores are both 7.  Since their SGI scores are the same and their carbon emission per capita are close, the only other determinant that is different in these countries are their GDP per capita’s.  The one for Italy is higher (37,227 USD in 2004) than the one in Spain (31,029 USD in 2005), however, they are still close to each other.  Therefore, in this case, we see only one year of difference in the starting year of the declining carbon curve.\\
\indent USA and Canada, on the other hand, have very high carbon emission per capita levels when compared to the most of the other high-income countries because petroleum and oil plays a major role in their economy.  The decreasing trend started in the USA in 1978 whereas it started in 1979 in Canada.  Their GDP per capita levels are around 29,000 USD level when this first sharp decrease started.  Then, there is another decrease point in Canada in 2006 when GDP per capita reached 45,858 and CO2 per capita level was at 17.56 metric tons per capita.  In the USA, there was also a second decrease trend in 2004 when the GDP per capita was at 47,288 USD  and carbon emission per capita was at 19.66 metric tons per capita.  In brief, in Canada and in the USA, the dependence of the economies on petroleum and oil interrupted the decreasing trends of their CO2 per capita curves for a period but, later, the decline continued nevertheless with the further increase in their GDP per capita levels. Furthermore, the level of CO2 per capita was still very high when compared to the other high-income countries in 2014, as shown in Figure 25 in the next section, due to the dependence in their economies on oil and petroleum. \\
\indent France and Belgium also had similar GDP per capita levels as the USA and Canada in 1979 when the decreasing trends on their CO2 per capita curves started.  GDP per capita of France was at 26,578 USD in 1979 whereas that of Belgium was at 25,946 USD.  The start of the declining trend of their carbon emission curves is the same for France and Belgium, 1974.  The carbon emission per capita for France (9.64 metric tons per capita) and Belgium (14.24 metric tons per capita) were lower than the ones for the USA and Canada at the start of the decreasing trend because the economies of France and Belgium are less dependent on oil and petroleum when compared to the USA and Canada.  Furthermore, the decreasing trend in France and Belgium were not interrupted like it was the case for the USA and Canada due to less dependence in their economies on oil and petroleum.  The SGI index for France is 7 whereas that of Belgium is 6.  Therefore, it can be said that since their GDP per capita levels were almost the same in 1979, the start of their carbon emission curves was 1979 for both of these countries despite the different scores on SGI. \\
\indent In brief, the EKC theory is valid for high-income countries that reached a certain country specific high level of GDP per capita.  When the carbon emission per capita vs GDP per capita of high-income countries are analyzed, it is clear that the GDP per capital is the main determinant for the carbon emission per capita trend for these selected countries.  It is also clear that the international institutions policies are not effective in starting the declining trend of the carbon emission per capita because the trend starts in a wide range of different years in these countries. The main factor for initiating the start of the decreasing trend is reaching the country specific high level of GDP per capita, rather than the international institutions policies. \\
\indent It is clear that reaching the country specific high level of GDP per capita is the main determinant for initiating the start of the declining trend of the carbon emission curve. The final level of carbon emission per capita that was reached and the time lag between the peak and the start of the declining curve are affected by a country’s economic dependence on petroleum/oil production and its SGI index score performance. If the economy is highly dependent on petroleum/oil/coal like Canada, the USA and Japan, the decreasing trend of carbon emission per capita is not as strong as the other high-income countries like Belgium, France and the UK.\\
\indent In summary, in this section, it is proven that the declining trend of the carbon emission trend starts only after the country specific high GDP per capita level is reached in these high-income countries.  Furthermore, the effects of the economic dependence on oil and petroleum and the performance on SGI score on carbon emission is also presented.
  \

\clearpage

\chapter{HIGH-INCOME COUNTRIES' CARBON EMISSION LEVELS }
\vspace*{0.4 cm}
\doublespacing
\indent In this section, the role of the government benevolence and the effect of the energy production dependency of the economy on carbon emission is analyzed. In Figure 25, the degradation levels as of 2014 for the selected high-income countries are presented.  The 2014 data is used in this analysis because the latest CO2 per capita data available for France and Italy is for 2014 in the World Bank’s database.

\begin{figure}[H]
	\centering
	\includegraphics[scale=0.8]{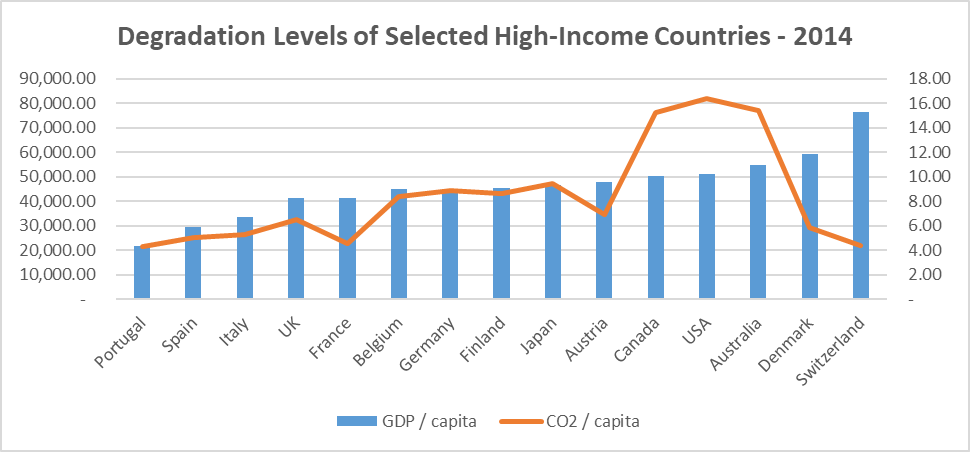}
		\caption{Fig 25:Degradation levels of countries in 2014}
\end{figure}

\indent In this graph, most of the countries’ carbon emission per capita levels are below 10 except for Canada, USA and Australia.  Even though the EKC theory is also valid for these exceptional countries, their carbon emission per capita levels are at significantly higher levels than the rest.  The reasons for these high levels of carbon emissions can be explained by their economic dependence on energy production and government benevolence.  SGI index shows the effectiveness of the environmental policies in countries. \\
\indent  The SGI index scores for Switzerland (9) and Denmark (9) are the highest.  Accordingly, these countries have relatively low carbon emission per capita emission levels as shown in Figure 27.  The SGI score of 9 shows that “Environmental policy goals are ambitious and effectively implemented as well as monitored within and across most relevant policy sectors that account for the largest share of resource use and emissions”(Stiftung, Bertelsmann 2020). As a result, it can be said that ambitious and effective environmental policies resulted in the relatively lower level of carbon emission per capita in Switzerland and Denmark. \\
\indent On the other hand, USA and Australia have lower SGI scores, 4 for both of them.  This shows that “Environmental policy goals are neither particularly ambitious nor are they effectively implemented and coordinated across relevant policy sectors” (Stiftung, Bertelsmann 2020). As a result, it can be stated that the insufficiency of the environmental policies result in USA’s and Australia’s high carbon emission per capita level. \\
\indent Canada, however, has a higher SGI score than USA and Australia.  Canada’s SGI score is 7 which means that “Environmental policy goals are mainly ambitious and effectively implemented and are monitored within and across some of the relevant policy sectors that account for the largest share of resource use and emissions.”(Stiftung, Bertelsmann 2020). Another important factor for Canada’s high carbon emission per capita level is the economy’s energy export dependency. Canada has to export energy to the USA (which means that they have to produce more energy than their citizens use) in order to maintain their economic wealth which results in a higher carbon emission per capita for the country when compared to the other selected high-income countries.  \\
\indent In summary, it is clear that the SGI score index of the country and the dependency of the economy on energy production with high carbon emission (coal, petroleum and oil) have a major role in the level of the CO2 per capita of the countries.

\chapter{CONCLUSION}
\vspace*{0.4 cm}
\doublespacing
\indent In this article, we analyze the carbon emission trends of twenty-four selected countries.  Our data analysis shows clearly that in all of the fifteen high-income countries that were selected, EKC hypothesis is valid and carbon emission starts to decline in high-income countries when a certain country specific high-income level is reached. On the other hand, it is too early to infer whether EKC hypothesis would be applicable to non-high income countries since their income levels are still at very low levels in the selected timeframe, which is from 1960-2016.  \\
 \indent Furthermore, our analysis showed clearly that the international policies were, unfortunately, ineffective in initiating a declining trend for carbon emission in high-income countries.  We also incorporated SGI scores in our analysis to assess the effect of the national efforts and policies on the start year of the carbon emission trend. \\  
 \indent In the last section, we focused on the carbon emission levels of all the selected high-income countries in 2014 and explained the relationship between their government benevolence represented by the SGI scores and their levels of carbon emission. Our analysis showed that there is a strong relationship between the overall SGI scores of these countries and their carbon emission levels. Furthermore, we also concluded that the dependence of the economy on energy production with high carbon emission (coal, petroleum and oil), also, have a major role in the level of the CO2 per capita of these countries.

\clearpage

\chapter*{REFERENCES}

\vspace{1cm}

 \singlespacing
\noindent Beer, Espen, and Prydz Divyanshi Wadhwa. “Classifying Countries by Income.\textit {The} \

\textit {World Bank IBRD IRA}, World Bank Group, 9   Sept. 2019, datatopics.worldbank.org/

\ world-development-indicators/stories/the-classification-of-countries-by-income.html.\
 
 \singlespacing
\noindent Callen, Tim, and Jonathan D. Ostry. “Japan's Lost Decade --- Policies for Economic \ 

Revival.” \textit {International Monetary Fund}, International Monetary Fund, 13 Feb. 

\ 2013, www.imf.org/external/pubs/nft/2003/japan/index.htm. 

\singlespacing
\noindent CO2 Emissions (Metric Tons per Capita).”  \textit {The World Bank IBRD IDA CO2 emissions} \ 

\textit {(metric tons per capita) Data},Carbon Dioxide Information Analysis \ 

\ Center, Environmental Sciences Division, Oak Ridge National Laboratory,

\ Tennessee, United States., 2021, data.worldbank.org/indicator/EN.ATM.CO2E.PC. 

\singlespacing 
\noindent Dasgupta, Susmita, et al. “Confronting the Environmental Kuznets Curve. \textit{Journal of} \

\textit{Economic Perspectives}, vol. 16, no. 1, 2002, pp. 147–168., doi:10.1257/0895330027157. 

\singlespacing 
\noindent “GDP per Capita (Constant 2010 USD).” textit{The World Bank IBRD IDA} \ 

\textit{GDP per Capita (Constant 2010 USD) | Data},World Bank National Accounts 

\ Data, OECD National Accounts Data Files, 2021, donnees.banquemondiale.org/

\singlespacing 
\noindent Grossman, Gene M., and Alan B. Krueger.“Environmental Impacts of a North \ 

American Free Trade Agreement.”  \textit{NBER}. \ National Bureau of Economic 

\ Research, Nov. 1991, www.nber.org/papers/w3914. \

\singlespacing
\noindent Hervieux, Marie-Sophie, and Pierre-Alexandre Mahieu. “A Detailed \ 

Systematic Review of the Recent Literature on Environmental Kuznets Curve 

\ Dealing with CO2.”  \textit{HAL Archives-Ouvertes.fr}, Laboratoire d’Economie 

\ Et De Management Nantes-Atlantique Université De Nantes , 2014,

\ hal.archives-ouvertes.fr/hal-01010243/document.

\singlespacing
\noindent Kuznets, Simon. “Economic Growth and Income Inequality.” \textit {The American} \

\textit {Economic Review}, XLV , no. 1, Mar. 1955, pp. 1–30.

\singlespacing
\noindent Stiftung, Bertelsmann. “SGI 2020: Downloads.” \textit{SGI 2020 | Downloads}, \

\ Sustainable Governance Indicators, 2020, www.sgi-network.org/2020/Downloads.

\singlespacing
\noindent US Energy Information Administration. \textit{U.S. Energy Information }\  

\textit{Administration - EIA - Independent Statistics and Analysis} 2 Dec. 2020, 

\ www.eia.gov/international/analysis/country/JPN.

\singlespacing
\noindent World Bank. “World Development Report 1992.” \textit {Open Knowledge Repository}, \ 

\ New York: Oxford University Press © World Bank, 1992, openknowledge.

\ worldbank.org/handle/10986/5975.

\end{document}